\documentclass[12pt,reqno]{amsart} 
\usepackage{xr}
\usepackage{amsthm}
\usepackage[mathscr]{eucal}
\usepackage{amsfonts}
\usepackage{amsmath}
\usepackage{amssymb}
\usepackage{graphicx}
\usepackage{xcolor}
\usepackage{pstricks}
\usepackage{fullpage}
\usepackage{hyperref}
\usepackage{booktabs}
\usepackage{colortbl}
\usepackage[foot]{amsaddr}
\usepackage{setspace}
\usepackage{tabularx}
\usepackage{natbib}
\usepackage{color}
\usepackage{tikz}

\usepackage{pdflscape}
\usepackage{floatrow}
\usepackage{subfig} 

\newcolumntype{N}{>{\centering\arraybackslash}m{1in}}

\newcolumntype{R}{>{\raggedright\arraybackslash}m{0.65in}}

\newcolumntype{W}{>{\centering\arraybackslash}m{0.9972in}}

\newcolumntype{Q}{>{\centering\arraybackslash}m{0.82in}}
\newcolumntype{P}{>{\raggedleft\arraybackslash}m{0.1025in}}

\newcolumntype{S}{>{\centering\arraybackslash}m{1.25in}}
\newcolumntype{T}{>{\centering\arraybackslash}m{0.5in}}
\newcolumntype{V}{>{\raggedleft\arraybackslash}m{0.9in}}

\newrgbcolor{gray70}{0.7019608 0.7019608 0.7019608}
\newrgbcolor{orangeR}{1 0.6470588 0}
\newrgbcolor{red4}{0.545098 0 0}
\newrgbcolor{olivedrab}{0.4196078 0.5568627 0.1372549}
\newrgbcolor{royalblue}{0.2549020 0.4117647 0.8823529}

\newcommand{\circleCol}[1]{\protect\tikz\protect\draw[#1, fill = #1] (0,0) circle (.5ex);}

\newcommand{\mb}[1]{\mathbb{#1}}
\newcommand{\mc}[1]{\mathcal{#1}}

\newcommand{\bs}[1]{\boldsymbol{#1}}

\newcommand{\msf}[1]{\mathsf{#1}}

\newcommand{\R}{\mb{R}}

\newcommand{\wh}[1]{\widehat{#1}}
\newcommand{\wt}[1]{\widetilde{#1}}

\newcommand{\vt}{\vartheta}

\newcommand{\argMax}{\msf{argmax}}
\newcommand{\SNR}{\msf{SNR}}
\newcommand{\TDE}{\msf{TDE}}

\newcommand{\Pn}{{\bf Pn}}

\newcommand{\PnCor}{{\bf Pn}_{\tt trim}}

\newcommand{\PnStd}{{\bf Pn}_{\tt standard}}

\newcommand{\LASSO}[1]{\textnormal{LASSO}^{#1}}

\newcommand{\LASSOCor}[1]{\textnormal{LASSO}_{\tt trim}^{#1}}

\newcommand{\LASSOCV}{\LASSO{{\tt CV}}} 
\newcommand{\LASSOCVCor}{\LASSO{\tt CV}_{\tt trim}} 

\title{Time delay estimation in satellite imagery time series 
of precipitation and NDVI: Pearson's cross correlation revisited}
\author{Inder Tecuapetla-G\'omez$^{(1)}$}
\email[Inder~Tecuapetla]{itecuapetla@conabio.gob.mx}
\address{$^1$CONACyT-CONABIO-Direcci\'on de Percepci\'on Remota\\
Liga Perif\'erico-Insurgentes Sur 4903\\
Parques del Pedregal, Tlalpan 14010, Ciudad de M\'exico}

\date{\today}

\begin{document}

\begin{abstract}
In order to describe more accurately the time relationships between daily satellite 
imagery time series of precipitation and NDVI we propose an estimator which
takes into account the \emph{sparsity} naturally observed in precipitation.
We conducted a series of simulation studies and show that the 
proposed estimator's variance is smaller than the canonical's (Pearson-based), 
in particular, when the signal-to-noise ratio is rather low. 
Also, the proposed estimator's variance was found smaller than the canonical's one
when we applied them to stacks of images (2002-2016) taken on some ecological 
regions of Mexico. 
Computations for this paper are based on functions
implemented in our new \texttt{R} package \texttt{geoTS} (available \href{https://github.com/inder-tg/geoTS}{here}).
\end{abstract}


\maketitle

\section{Introduction and Dataset}~\label{sec.intro}


{T}{ime} {delay estimation} consists of
approximating the apparent shift between an emmited signal (also known 
as reference signal) and another one which is received in a different point 
in time and space (delayed signal).

Estimating this apparent delay between general signals has been a subject 
of study in fields as diverse as time series (\cite{Hamon.Hannan.1974}, \cite{Hannan.1988}), 
signal processing (\cite{Knapp.Carter.76}, \cite{Carter.1981}),
radar, sonar and seismology \cite{Benesty.2004},
and remote sensing. Indeed, considering time series of satellite imagery as 
discretized signals, \cite{vicente2013}, \cite{barbosa2015} and \cite{Colditz.etal.2017},
among others, have conducted global and regional studies seeking to determine 
the time relations between precipitation and vegetation growth (described, 
e.g.~by the NDVI, which will also be used here).
Naturally, precipitation is a \emph{sparse} variable, that is, it may have 
prolonged periods of dry days (zero precipitation), and this characteristic has not been
taken into consideration in the aforementioned studies. 

In this paper we revisit the canonical tool to estimate the delay between discrete
signals, the Pearson's product moment association 
and propose to incorporate the precipitation' sparsity, through solving a LASSO regression problem, in determining
the delay. The resulting time delay estimator ($\TDE$) has proven to reduce the variance
of the canonical estimator in many cases, see Section~\ref{sec.sims} for 
several simulation studies.

\begin{figure}[htb]
    \centering
        \scalebox{.9}{\sidesubfloat[]{\includegraphics[width=0.35\linewidth]{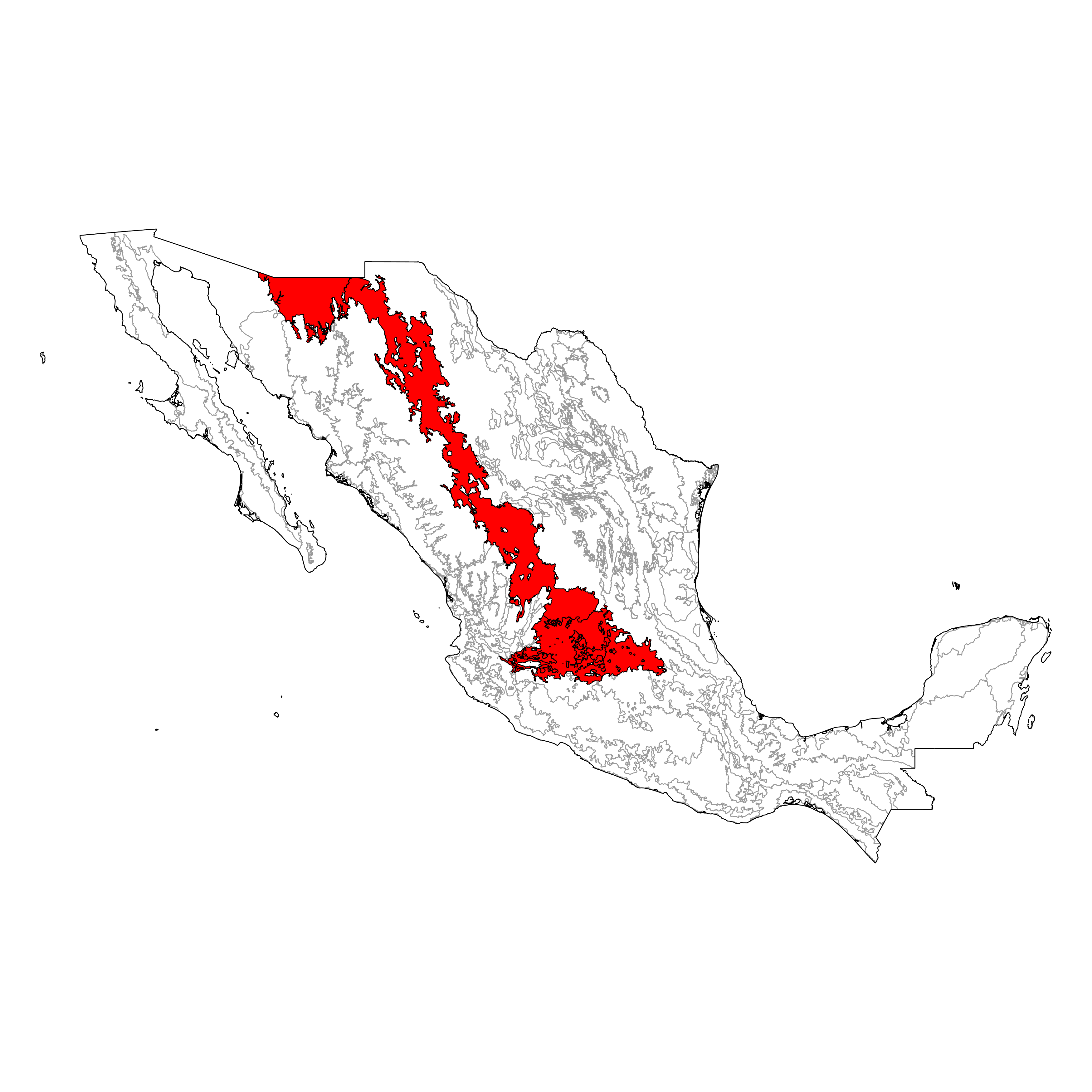}\label{Fig1A}}}
        \scalebox{.8}{\sidesubfloat[\vspace{-5cm}]{\includegraphics[width=0.35\linewidth]{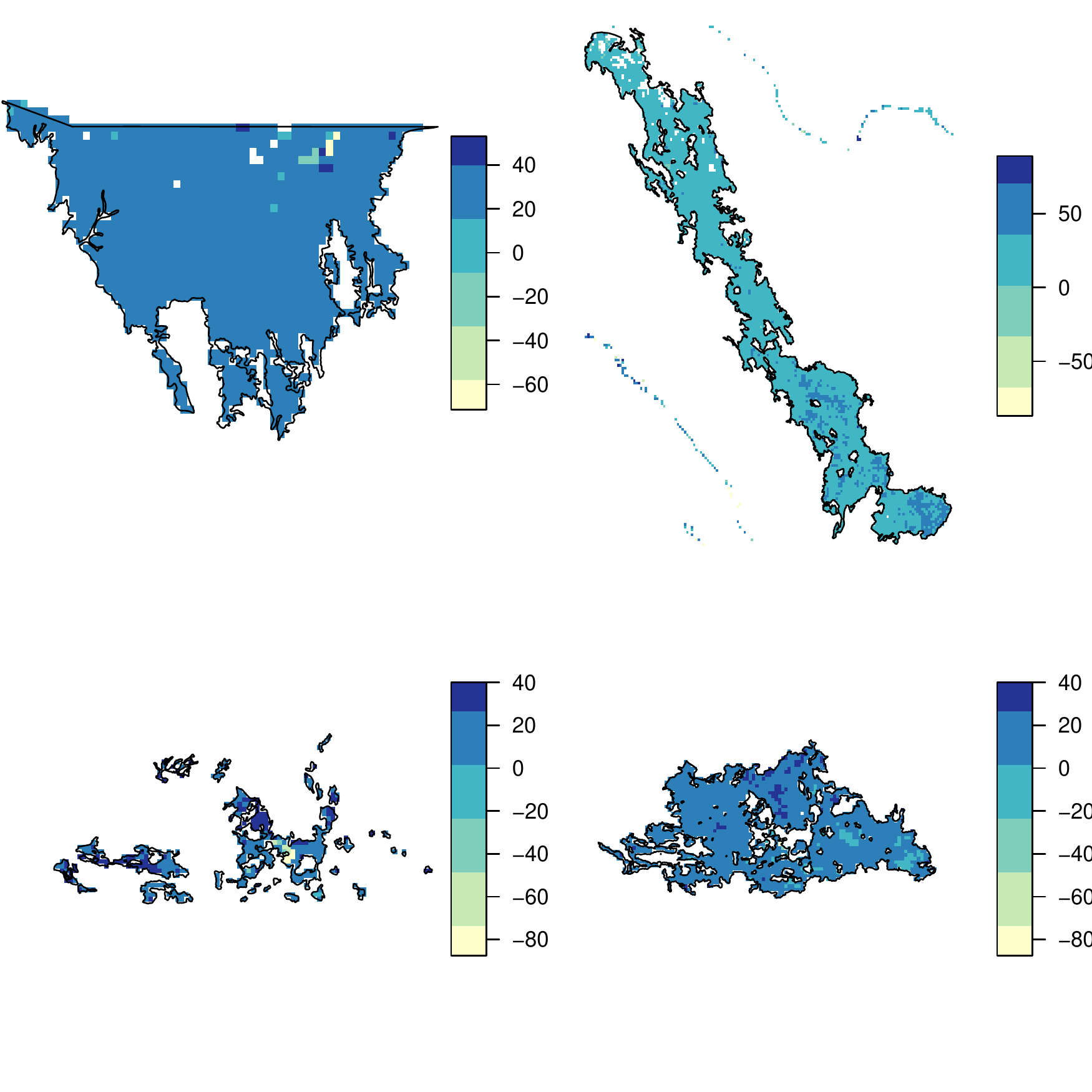}\label{Fig1B}}}%
        \scalebox{.8}{\sidesubfloat[\vspace{-5cm}]{\includegraphics[width=0.35\linewidth]{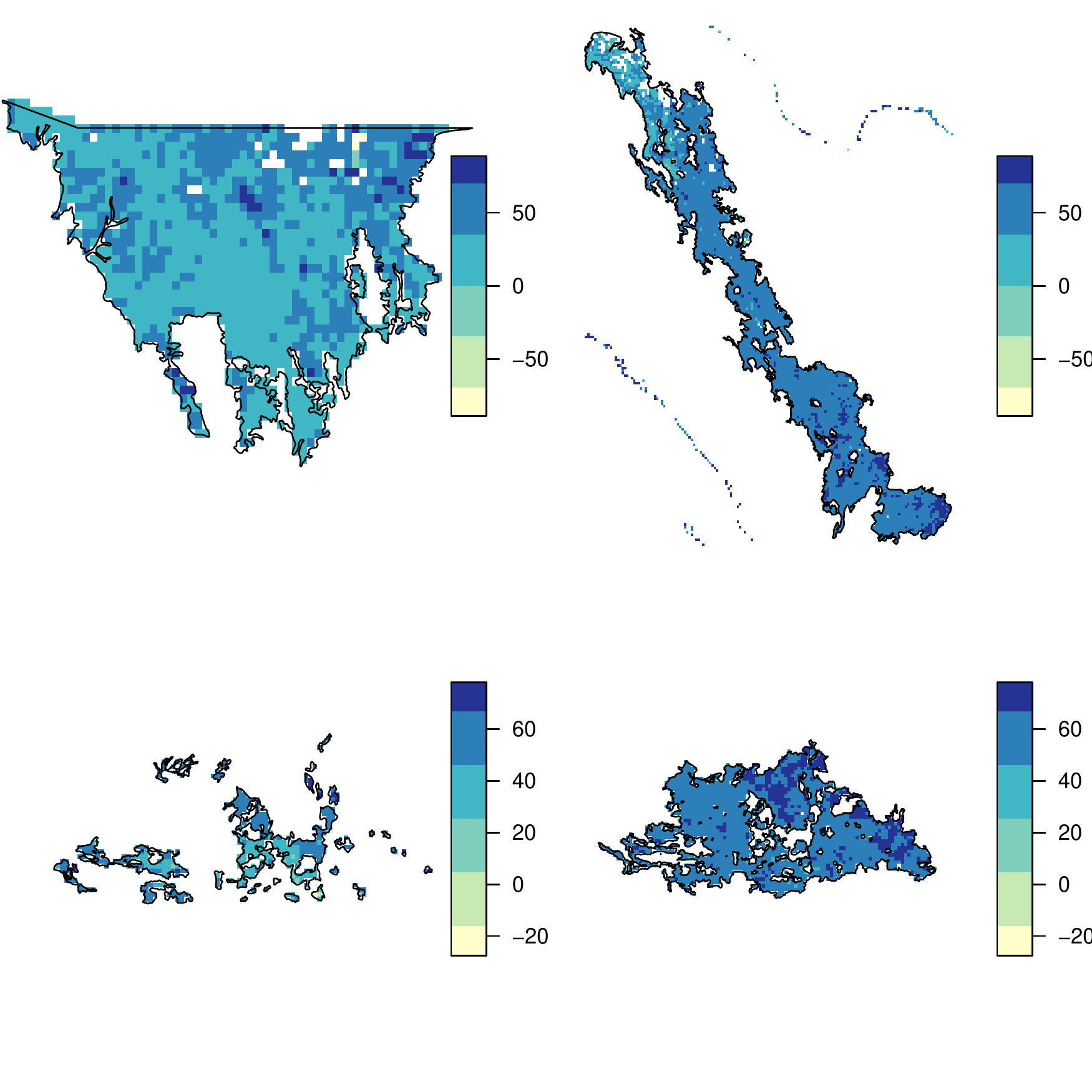}\label{Fig1C}}}

        \vspace{-1cm}
        \caption{\footnotesize {
        (A) Semiarid Southern Uplands eco-region (red).
        (B) Median of significant (at 5\%) $\LASSOCVCor$ time delay estimates obtained from
        satellite imagery time series of daily precipitation and NDVI from 2002 to 2016 in regions 
        (highlighted in (A), from top-left to bottom-right) Madrense, Mezquital, 
        Interior Plains and Plateau Plains.
        (C) Median of significant (at 5\%) $\Pn$ time delay estimates obtained from same products, time and regions described
        in (B).
        }}      
\end{figure}

We have applied our approach to daily time series of precipitation and NDVI (from 2002 to 2016)
taken on the Semi-arid Southern Uplands eco-region located in the Mexican
Territory, see Fig.\protect\ref{Fig1A}. This eco-region is comprised by areas
whose precipitation and vegetation show similar dynamics although they may include more than one ecosystem, 
cf.~Ch.3~p.104 of \cite{CONABIO.2008}.
This feature adds variability to any $\TDE$ and to the best of our knowledge
an assessment of this variability still remains elusive and is one of the contributions 
of this work, see Section~\ref{sec.apps}.




We utilize the Climate Hazards Group InfraRed Precipitation with Station (CHIRPS) dataset (ver.~2.0),
see \cite{Funk.2015}. The NDVI used in this paper was derived
from NASA's product MCD43C4 (ver.6), cf.~\cite{Schaaf.Wang.2015}. The spatial resolution
of these two products is $0.05^\circ$. We re-sampled the eco-region shapefile,
available online~\cite{ecoRegions.2007}, so that its spatial resolution coincide with the other two.


Notation: An $l$-dimensional vector full with zeros will be denoted as
$\bs{0}_{l}$; for a generic vector we write ${\bs v}_{[i:j]} = \begin{pmatrix} v_i,\ldots, v_j \end{pmatrix}^\top$, 
$i < j$.

\floatsetup[figure]{style=plain, subcapbesideposition = top}

\section{Methodology}~\label{sec.methods}
Let us introduce
the time delay estimation problem based on the Pearson's product moment
association function. Let ${\bs x} = (x_1,\ldots,x_n)^\top$ and ${\bs y} = (y_1,\ldots,y_n)^\top$
denote the time series of precipitation and vegetation index, respectively; $n$ 
denotes the sample size. Let $\mc G_n$ denote the grid of $2n-1$ points 
$\{-(n-1), \ldots, -1, 0, 1, \ldots, n-1\}$. For $l\in \mc G_n$ the general 
Pearson's product moment association between $\bs{x}$ and $\bs{y}$ is defined as
\begin{equation}~\label{eq.PearsonFun}
    \gamma_l 
    = 
    \begin{cases}
     \frac{1}{n-|l|}\,\sum_{k=1}^{n-|l|}\,M(x_{|l|+k}; {\bs x})\,M(y_k; {\bs y}), & l < 0\\
     \frac{1}{n-l}\,\sum_{k=1}^{n-l}\,M(x_k; {\bs x})\,M(y_{k+l}; {\bs y}), & l \geq 0
    \end{cases},
\end{equation}
where $M(x_k; \bs{x})$ can be equal to either $x_k$ (unscaled),
$(x_k - \bar{\bs{x}})/\wh{\sigma}_{\bs{x}}$ (standard) or 
$(x_k - \bar{\bs{x}}_{trim})/\wh{\sigma}_{\bs{x}}^{trim}$ (trimmed).
Above, $\bar{\bs x}$ and $\wh{\sigma}_{\bs x}$ are the arithmetic mean and 
a standard deviation (s.d.) estimate of ${\bs x}$. Similarly, $\bar{\bs x}_{trim}$
and $\wh{\sigma}_{\bs x}^{trim}$ denote mean and s.d., 
respectively, of $\bs{x}_{[(|l|+1):n]}$ for $l < 0$ or $\bs{x}_{[1:(n-l)]}$ for $l \geq 0$.
This generality in the computation of the empirical product
mean between ${\bs x}$ and ${\bs y}$ allows us to study
Pearson's correlation and other popular forms of 
association considered in the literature of the time delay problem.
In what follows we will assume that ${\bs x}$ and ${\bs y}$ have been standardized.

Let $\bs{\Gamma} = \begin{pmatrix} \gamma_{-(n-1)}, \ldots, 
               \gamma_{-1}, \gamma_0, \gamma_{1},\ldots,
               \gamma_{n-1} \end{pmatrix}^\top$.
Note that \eqref{eq.PearsonFun} can be written in a matricial form.
Indeed,
\begin{equation}~\label{eq.PearsonMatricial}
    \bs{\mc S}^\top\,\bs{x} = \bs{J}_n\,\bs{\Gamma}, 
\end{equation}
where $\bs{J}_n$ is a $2n-1$ squared diagonal matrix with main diagonal equal to
$(1, 2, \ldots, n-1, n, n-1, \ldots, 2, 1)$ and $\bs{\mc S}\in \R^{n\times(2n-1)}$
is the \emph{sparse} matrix whose $i$-th column is,
\[    
    \bs{\mc S}[\cdot, i] 
        =
    \begin{cases}
        \bs{a}_{i,n}^\top & 1 \leq i \leq n,\\
        \bs{b}_{i,n}^\top & n+1 \leq i \leq 2n-1
    \end{cases},
\]
$\bs{a}_{i,n}=(\bs{0}_{n-i}, \bs{y}_{[1:i]})$ and $\bs{b}_{i,n}=(\bs{y}_{[(i+1-n):n]}, \bs{0}_{i-n})$.
The $\TDE$ between ${\bs x}$ and ${\bs y}$ is by definition,
\begin{equation}~\label{eq.TiDE}
 \wh{l}_n 
 = 
 \argMax_{l\in \mc G_n}\, {\gamma}_l^2. 
\end{equation}

In the last two decades the least absolute shrinkage and selection operator (LASSO) has
been applied to myriads of practical problems in a wide range of the sciences, cf.~\cite{Tibshirani.2011}.
Now we argue that this operator can be used in the context of time delay estimation and,
as it is shown in our simulations, the use of LASSO improves the precision
of the resulting $\TDE$. In a nutshell, LASSO is regression
with an $\ell_1$-norm penalty and this constraint induces \emph{sparsity} in the estimated
coefficients of the regression, i.e., the LASSO solution is
a vector with a fair amount of entries which are identically zero. Although
${\bs x}$ is already a sparse vector, as it contains measurements of a precipitation
process, 
we argue that 
by fitting a LASSO regression we can write ${\bs x}$ in a far more sparse 
representation and this will ease the solution of \eqref{eq.TiDE}.
Namely, we assume that there exists a highly sparse vector ${\bs s}\in \R^{2n-1}$, such that
\[
    \bs{x} = \bs{\mc S}{\bs s} + {\bs \varepsilon},
\]
where ${\bs \varepsilon} = \begin{pmatrix} \varepsilon_1,\ldots,\varepsilon_n \end{pmatrix}^\top$ 
contains error terms (possibly correlated). Then, we seek a vector $\bs{s}_{\ell_1}\in \R^{2n-1}$, 
which minimizes the penalized least squares problem
\begin{equation}\label{eq.Bridge}
    \sum_{i=1}^n( x_i - {\bs{\mc S}}[i,\cdot] \bs{s} )^2 + \lambda_n \sum_{j=1}^{2n-1}|s_j|, 
\end{equation}
for a given tuning parameter $\lambda_n$. Having determined $\bs{s}_{\ell_1}$, let 
$\wt{\bs x} = \bs{\mc S}\bs{s}_{\ell_1}$ and plug it into 
\eqref{eq.PearsonFun} to produce a new $\TDE$. 
That is, in this paper we propose 
\begin{equation}~\label{eq.sTiDE}
\wt{l}_n 
=
\argMax_{l\in \mc G_n}\, {\gamma}_l^2(\wt{\bs x}, {\bs y})
\end{equation}
as an estimator which takes into account the sparse nature of the original 
precipitation process ${\bs x}$. 

\subsection{On the penalization parameter}

In order to solve \eqref{eq.Bridge} we need to provide a tuning parameter $\lambda_n$. 
Note that the larger the value of $\lambda_n$ the more
sparsity we impose in the solution $\bs{s}_{\ell_1}$ so that this parameter must be carefully
selected to avoid an overpenalization leading to meaningless estimators, e.g. the vector zero as 
the new representation of ${\bs x}$.

The standard algorithm to solve \eqref{eq.Bridge} provides an entire \emph{solution path} 
(a set of values for $\lambda_n$ and its corresponding LASSO estimate), cf.~\cite{efron2004}. 
As \emph{leave-one-out} 
cross validation produces a tuning parameter whose LASSO estimate 
achieves the least predictive error we suggest its use to select $\lambda_n$.
Another, less time consuming, empirical strategies will be discussed in Section~\ref{sec.sims}.

\subsection{On the search grid}

We introduced problems \eqref{eq.TiDE}-\eqref{eq.sTiDE} considering the 
general grid $\mc G_n$ but it is not recommended to optimize $\gamma_l^2$ over the entire 
grid.
Indeed, note that at the left and right extremes of $\mc G_n$, $\gamma_l^2$ exhibits 
a great fluctuation as at those extremes only a small amount of observations are used to
calculate the association between ${\bs x}$ and ${\bs y}$.
Therefore it is common to utilize a smaller grid to solve \eqref{eq.TiDE} or \eqref{eq.sTiDE}.

In studies involved in assessing the time relationships between daily precipitation 
and vegetation growth, over a year, a grid of $\pm 3$ months has been used, 
e.g.~\cite{vicente2013} and \cite{barbosa2015}; 
effectively, this represents to using only 25\% of $\mc G_n$. 
In our applications we used grids of $\pm 2$, $\pm 3$ and $\pm 4$ months without
finding major differences between the estimates. See Section~\ref{subsec.Results}
for comments on the relationship of signal-to-noise ratio ($\SNR$) and grid length.


\section{Simulations}~\label{sec.sims}

In this section we assess the performance of the $\TDE$s proposed in Section~\ref{sec.methods}. 
Now we introduce notation for each of the methods under comparison.

$\bs{\vt}_{\tt option}^{\tt type}$ denotes a $\TDE$ in which the association
function and tuning parameter are settled by ${\tt option}$ and ${\tt type}$, respectively.
We used $\bs{\vt} = (\Pn, \LASSO{})$, i.e., a Pearson-based (via \eqref{eq.TiDE}) or a
LASSO-based (via \eqref{eq.sTiDE}) estimator. 
For ${\tt option}$ we consider
\texttt{trim} and \texttt{standard}, see \eqref{eq.PearsonFun}; the \texttt{unscaled} case
is also consider (we simply leave a blank space). When ${\tt type} = 0.1$ 
we selected the $10\%$ quantile of the tuning parameter ($\lambda_n$) distribution (a very fast procedure)
and when ${\tt type = CV}$ we used $\lambda_n$ provided by cross validation.
Each simulation setup will be repeated 1,000 times and we assess the behavior of these estimators
through their mean and s.d.


\subsection{Simulation of the precipitation process}~\label{subsec.PrecProcess}

In order to simulate easily a realistic daily precipitation time series of arbitrary length
we use an \emph{occurrence and amount} model.
We simulate occurrences through a discrete Markov chain model with 2 states (dry day / wet day).
In order to implement this model we need estimates of matrices of transition probabilities; 
further details are given below. We simulate rainfall amount from an exponentially distributed
random variable (r.v.) as this is known to be appropriate and allows us 
to assess the performance of $\TDE$s as a function of the $\SNR$.
Now we explain how we estimate the necessary parameters to execute our simulations.

\begin{table*}[htb]
\centering
\scalebox{0.55}{
\begin{tabular}{R WWW P WWW P WWW} 
  \toprule[1.25pt]
      
  $\TDE$s & \multicolumn{3}{c}{$\tau=37$} && \multicolumn{3}{c}{$\tau=110$} && \multicolumn{3}{c}{$\tau=183$} \\ 
          & \multicolumn{3}{c}{Exp($\lambda$) errors} && \multicolumn{3}{c}{Exp($\lambda$) errors} && \multicolumn{3}{c}{Exp($\lambda$) errors}\\
    
  \midrule[1.25pt] 
  & \multicolumn{1}{c}{$\lambda = 0.125$} & $\lambda = 0.5$ & $\lambda = 2.5$ 
  &
  & \multicolumn{1}{c}{$\lambda = 0.125$} & $\lambda = 0.5$ & $\lambda = 2.5$ 
  &
  & \multicolumn{1}{c}{$\lambda = 0.125$} & $\lambda = 0.5$ & $\lambda = 2.5$ \\
  \cmidrule[1.25pt]{2-4}
  \cmidrule[1.25pt]{6-8}
  \cmidrule[1.25pt]{10-12} 
  
  
  \multicolumn{12}{c}{Madrense}\\
  
  $\Pn$ & {\footnotesize 37.001 (0.032)} & {\footnotesize 37.108 (0.524)} & {\footnotesize 38.55 (3.516)} && 
  {\footnotesize 110.001 (0.032)} & {\footnotesize 110.045 (0.315)} & {\footnotesize 110.903 (2.164)} && 
  {\footnotesize 183 (0)} & {\footnotesize 182.998 (0.045)} & {\footnotesize 182.981 (0.25)} \\ 
  $\PnCor$ & {\footnotesize 37 (0)} & {\footnotesize 38.01 (10.3)} & {\footnotesize 40.97 (18.586)} && 
  {\footnotesize 110 (0)} & {\footnotesize 110.001 (0.152)} & {\footnotesize 110.038 (1.335)} && 
  {\footnotesize 183 (0)} & {\footnotesize 182.992 (0.089)} & {\footnotesize 181.695 (18.586)} \\ 
  $\PnStd$ & {\footnotesize 37 (0)} & {\footnotesize 37.062 (0.395)} & {\footnotesize 38.068 (5.473)} && 
  {\footnotesize 110 (0)} & {\footnotesize 110.02 (0.199)} & {\footnotesize 110.142 (1.396)} && 
  {\footnotesize 183 (0)} & {\footnotesize 182.996 (0.063)} & {\footnotesize 182.944 (0.464)} \\ 
  $\LASSO{0.1}$ & {\footnotesize 37 (0)} & {\footnotesize 37.047 (0.345)} & {\footnotesize 38.532 (3.504)} && 
  {\footnotesize 110 (0)} & {\footnotesize 110.019 (0.197)} & {\footnotesize 110.886 (2.098)} && 
  {\footnotesize 183 (0)} & {\footnotesize 182.996 (0.063)} & {\footnotesize 182.982 (0.248)} \\ 
  $\LASSOCor{0.1}$ & {\footnotesize 37 (0)} & {\footnotesize 37.012 (0.189)} & {\footnotesize 50.383 (34.935)} && 
  {\footnotesize 110 (0)} & {\footnotesize 109.993 (0.114)} & {\footnotesize 110.01 (1.177)} && 
  {\footnotesize 183 (0)} & {\footnotesize 182.992 (0.089)} & {\footnotesize 182.882 (0.876)} \\ 
  $\LASSOCV$ & {\footnotesize 37 (0)} & {\footnotesize 37.099 (0.497)} & {\footnotesize 38.405 (3.374)} && 
  {\footnotesize 110.001 (0.032)} & {\footnotesize 110.045 (0.315)} & {\footnotesize 110.811 (2.085)} && 
  {\footnotesize 183 (0)} & {\footnotesize 182.998 (0.045)} & {\footnotesize 182.978 (0.279)} \\ 
  $\LASSOCVCor$ & {\footnotesize 37 (0)} & {\footnotesize 37.342 (5.967)} & {\footnotesize 39.505 (14.78)} && 
  {\footnotesize 110 (0)} & {\footnotesize 109.993 (0.114)} & {\footnotesize 109.986 (1.072)} && 
  {\footnotesize 183 (0)} & {\footnotesize 182.993 (0.083)} & {\footnotesize 182.89 (0.82)} \\ 
  
  \multicolumn{12}{c}{Mezquital}\\
  
  $\Pn$ & {\footnotesize {\footnotesize 37 (0)}} & {\footnotesize {\footnotesize 37.007 (0.217)}} & {\footnotesize {\footnotesize 37.469 (3.58)}} && {\footnotesize {\footnotesize 110.003 (0.055)}} & {\footnotesize {\footnotesize 110.171 (0.671)}} & {\footnotesize {\footnotesize 111.499 (3.366)}} && {\footnotesize {\footnotesize 183 (0)}} & {\footnotesize {\footnotesize 182.998 (0.045)}} & {\footnotesize {\footnotesize 182.932 (0.337)}} \\ 
  $\PnCor$ & {\footnotesize {\footnotesize 37 (0)}} & {\footnotesize {\footnotesize 41.772 (22.373)}} & {\footnotesize {\footnotesize 48.541 (33.669)}} && {\footnotesize {\footnotesize 110 (0)}} & {\footnotesize {\footnotesize 110.02 (0.286)}} & {\footnotesize {\footnotesize 110.394 (2.444)}} && {\footnotesize {\footnotesize 183 (0)}} & {\footnotesize {\footnotesize 182.97 (0.182)}} & {\footnotesize {\footnotesize 182.079 (11.197)}} \\ 
  $\PnStd$ & {\footnotesize {\footnotesize 37 (0)}} & {\footnotesize {\footnotesize 36.983 (0.207)}} & {\footnotesize {\footnotesize 39.332 (16.903)}} && {\footnotesize {\footnotesize 110 (0)}} & {\footnotesize {\footnotesize 110.054 (0.333)}} & {\footnotesize {\footnotesize 110.588 (2.466)}} && {\footnotesize {\footnotesize 183 (0)}} & {\footnotesize {\footnotesize 182.991 (0.094)}} & {\footnotesize {\footnotesize 182.829 (0.736)}} \\ 
  $\LASSO{0.1}$ & {\footnotesize {\footnotesize 37 (0)}} & {\footnotesize {\footnotesize 36.985 (0.225)}} & {\footnotesize {\footnotesize 37.462 (3.573)}}&&
  {\footnotesize {\footnotesize 110 (0)}} & {\footnotesize {\footnotesize 110.045 (0.295)}} & {\footnotesize {\footnotesize 111.489 (3.301)}} && {\footnotesize {\footnotesize 183 (0)}} & {\footnotesize {\footnotesize 182.994 (0.077)}} & {\footnotesize {\footnotesize 182.936 (0.341)}} \\ 
  $\LASSOCor{0.1}$ & {\footnotesize {\footnotesize 37 (0)}} & {\footnotesize {\footnotesize 36.971 (0.224)}} & 
  {\footnotesize {\footnotesize 67.231 (48.797)}} && 
  {\footnotesize {\footnotesize 110 (0)}} & {\footnotesize {\footnotesize 110.003 (0.239)}} & {\footnotesize {\footnotesize 110.179 (2.025)}} && {\footnotesize {\footnotesize 183 (0)}} & {\footnotesize {\footnotesize 182.973 (0.174)}} & {\footnotesize {\footnotesize 182.711 (1.098)}} \\ 
  $\LASSOCV$ & {\footnotesize {\footnotesize 37 (0)}} & {\footnotesize {\footnotesize 37.005 (0.207)}} & 
  {\footnotesize {\footnotesize 37.329 (2.878)}} && 
  {\footnotesize {\footnotesize 110.003 (0.055)}} & {\footnotesize {\footnotesize 110.161 (0.646)}} & {\footnotesize {\footnotesize 111.242 (3.133)}} && {\footnotesize {\footnotesize 183 (0)}} & {\footnotesize {\footnotesize 182.997 (0.055)}} & {\footnotesize {\footnotesize 182.933 (0.347)}} \\ 
  $\LASSOCVCor$ & {\footnotesize {\footnotesize 37 (0)}} & {\footnotesize {\footnotesize 37.515 (7.683)}} & 
  {\footnotesize {\footnotesize 45.953 (30.083)}} && 
  {\footnotesize {\footnotesize 110 (0)}} & {\footnotesize {\footnotesize 110.007 (0.226)}} & {\footnotesize {\footnotesize 110.036 (1.68)}} && {\footnotesize {\footnotesize 183 (0)}} & {\footnotesize {\footnotesize 182.975 (0.162)}} & {\footnotesize {\footnotesize 182.719 (1.064)}} \\ 
  
  \multicolumn{12}{c}{Interior Plains}\\
  
  $\Pn$ & {\footnotesize 37 (0)} & {\footnotesize 37.056 (0.37)} & {\footnotesize 38.11 (5.152)} && 
  {\footnotesize 110.001 (0.032)} & {\footnotesize 110.124 (0.691)} & {\footnotesize 111.282 (3.852)} && 
  {\footnotesize 183 (0)} & {\footnotesize 182.992 (0.089)} & {\footnotesize 182.803 (0.885)} \\ 
  $\PnCor$ & {\footnotesize 37 (0)} & {\footnotesize 44.61 (27.828)} & {\footnotesize 54.721 (42.69)} && 
  {\footnotesize 110 (0)} & {\footnotesize 109.968 (0.419)} & {\footnotesize 107.534 (19.156)} && 
  {\footnotesize 182.999 (0.032)} & {\footnotesize 182.919 (0.423)} & {\footnotesize 177.286 (31.137)} \\ 
  $\PnStd$ & {\footnotesize 37 (0)} & {\footnotesize 37.004 (0.276)} & {\footnotesize 39.63 (24.871)} && 
  {\footnotesize 110 (0)} & {\footnotesize 110.007 (0.315)} & {\footnotesize 109.849 (2.701)} && 
  {\footnotesize 182.999 (0.032)} & {\footnotesize 182.976 (0.172)} & {\footnotesize 182.46 (1.785)} \\ 
  $\LASSO{0.1}$ & {\footnotesize 37 (0)} & {\footnotesize 37.007 (0.239)} & {\footnotesize 38.041 (4.901)} && 
  {\footnotesize 110 (0)} & {\footnotesize 110.016 (0.303)} & {\footnotesize 111.276 (3.813)} && 
  {\footnotesize 182.999 (0.032)} & {\footnotesize 182.983 (0.144)} & {\footnotesize 182.829 (0.861)} \\ 
  $\LASSOCor{0.1}$ & {\footnotesize 37 (0)} & {\footnotesize 36.966 (0.348)} & {\footnotesize 70.346 (56.706)} && 
  {\footnotesize 110 (0)} & {\footnotesize 109.957 (0.368)} & {\footnotesize 108.045 (16.355)} && 
  {\footnotesize 182.999 (0.032)} & {\footnotesize 182.94 (0.347)} & {\footnotesize 181.412 (9.817)} \\ 
  $\LASSOCV$ & {\footnotesize 37 (0)} & {\footnotesize 37.048 (0.352)} & {\footnotesize 37.854 (4.757)} && 
  {\footnotesize 110 (0)} & {\footnotesize 110.116 (0.667)} & {\footnotesize 110.944 (3.406)} && 
  {\footnotesize 183 (0)} & {\footnotesize 182.991 (0.105)} & {\footnotesize 182.818 (0.88)} \\ 
  $\LASSOCVCor$ & {\footnotesize 37 (0)} & {\footnotesize 37.738 (9.088)} & {\footnotesize 50.527 (37.889)} && 
  {\footnotesize 110 (0)} & {\footnotesize 109.96 (0.356)} & {\footnotesize 109.418 (2.476)} && 
  {\footnotesize 182.999 (0.032)} & {\footnotesize 182.942 (0.345)} & {\footnotesize 181.809 (7.141)} \\ 

  \multicolumn{12}{c}{Plateau Plains}\\
  
  $\Pn$ & {\footnotesize 37 (0)} & {\footnotesize 37.051 (0.665)} & {\footnotesize 37.655 (4.619)} && 
  {\footnotesize 110.001 (0.032)} & {\footnotesize 110.122 (0.536)} & {\footnotesize 111.345 (3.529)} && 
  {\footnotesize 183 (0)} & {\footnotesize 182.996 (0.063)} & {\footnotesize 182.883 (0.984)} \\ 
  $\PnCor$ & {\footnotesize 37 (0)} & {\footnotesize 42.109 (23.085)} & {\footnotesize 47.679 (34.075)} && 
  {\footnotesize 110 (0)} & {\footnotesize 110.002 (0.279)} & {\footnotesize 110.271 (3.008)} && 
  {\footnotesize 183 (0)} & {\footnotesize 182.96 (0.297)} & {\footnotesize 182.196 (6.279)} \\ 
  $\PnStd$ & {\footnotesize 37 (0)} & {\footnotesize 37.003 (0.247)} & {\footnotesize 39.885 (20.761)} && 
  {\footnotesize 110 (0)} & {\footnotesize 110.022 (0.286)} & {\footnotesize 110.481 (2.905)} && 
  {\footnotesize 183 (0)} & {\footnotesize 182.982 (0.209)} & {\footnotesize 182.761 (1.468)} \\ 
  $\LASSO{0.1}$ & {\footnotesize 37 (0)} & {\footnotesize 37.007 (0.239)} & {\footnotesize 37.615 (4.562)} && 
  {\footnotesize 110 (0)} & {\footnotesize 110.028 (0.299)} & {\footnotesize 111.355 (3.57)} && 
  {\footnotesize 183 (0)} & {\footnotesize 182.991 (0.094)} & {\footnotesize 182.897 (0.973)} \\ 
  $\LASSOCor{0.1}$ & {\footnotesize 37 (0)} & {\footnotesize 36.975 (0.273)} & {\footnotesize 69.22 (51.076)} && 
  {\footnotesize 110 (0)} & {\footnotesize 109.989 (0.23)} & {\footnotesize 109.971 (6.053)} && 
  {\footnotesize 183 (0)} & {\footnotesize 182.965 (0.264)} & {\footnotesize 182.627 (1.71)} \\ 
  $\LASSOCV$ & {\footnotesize 37 (0)} & {\footnotesize 37.047 (0.661)} & {\footnotesize 37.548 (4.43)} && 
  {\footnotesize 110.001 (0.032)} & {\footnotesize 110.105 (0.496)} & {\footnotesize 111.105 (3.289)} && 
  {\footnotesize 183 (0)} & {\footnotesize 182.996 (0.063)} & {\footnotesize 182.896 (0.974)} \\ 
  $\LASSOCVCor$ & {\footnotesize 37 (0)} & {\footnotesize 37.416 (6.89)} & {\footnotesize 46.662 (31.429)} && 
  {\footnotesize 110 (0)} & {\footnotesize 109.992 (0.228)} & {\footnotesize 109.854 (6.005)} && 
  {\footnotesize 183 (0)} & {\footnotesize 182.966 (0.263)} & {\footnotesize 182.645 (1.636)} \\ 

  \bottomrule[1.25pt]
\end{tabular}}
\caption{\small{Mean and s.d.~(in parenthesis).}~ \label{tab_mean_sd_ExpRainfall}}
\end{table*}

\subsection{Parameters to simulate precipitation processes by eco-region}~\label{subsec.PrecProcessEcoRegions}

We utilize the characteristics of occurrences and rainfall amount from the time
series of precipitation taken on the eco-region shown in Fig.\protect\ref{Fig1A} as 
this datacube will be further analyzed in Section~\ref{sec.apps}.

We utilize the 2004 images as this stack has the best quality data.
We have grouped the pixels of these images by region and 
fitted an exponential density to each monthly precipitation distribution.
Overall, the exponential fit seems to be appropriate for low precipitation 
whereas tends to underestimate slightly the remaining part of the distribution. 

We use each region's pixels to calculate the corresponding transition probability matrix. 
At a given pixel, we compute the frequency
at which the precipitation time series transitions from a dry day (zero precipitation)
to a wet day (nonzero precipitation); a similar calculation is done to account for the 
events (dry/dry), (wet/dry) and (wet/wet). 

\subsection{Basic impulse and exponentially distributed rainfall}~\label{subsec.ExpRainfall}

We found the use of basic \emph{impulse} functions as an effective tool to generate signals
and control the apparent delay between them.
An impulse function will take the value zero everywhere
except in a finite interval where the function will equal the unity. 
For example, let $\tau \geq 0$ and define $f(t)$ equal to zero for $t\in (-\infty,110)\cup[183, 366]$
and one for $t\in [110,183)$; also set $g(t) = f(t-\tau)$,
note that $g$ is in essence $f$ when the latter function \emph{has walked} 
$\tau$ time-points forward.
That is, the time delay between $f$ and $g$ is the (known) parameter $\tau$.
In one of our simulations we consider the functions just introduced 
with $\tau=37$; in other simulations we used $\tau = 110, 183$ with changes in the
end-points of the finite interval where $f$ and $g$ are nonzero.

The next step consists of superimposing the realization of the precipitation process
described above to the function $f$ whereas the function $g$ is perturbed by a normal r.v.
with a small s.d.; we use $\sigma_d=0.0075$ throughout all our simulations
as this value is in line with the variability of the NDVI analyzed in Section~\ref{sec.apps}.
As a result we end up with one time series with high variance (the superimposition of occurrences 
on $f$)
and another with considerably lower variance ($g+\sigma_d^2$);
this seeks to emulate the different levels of variability between a (real)
time series of precipitation and its vegetation index counterpart.

The superimposition is done as follows.
If $t_\ast$ corresponds to a wet day 
we set $f(t_\ast)=1$. Observe that then $f$ is nonzero in the interval $[110, 183)$ and in all 
those $t_\ast$ timepoints. Next we perturb $f$, additively, at each point where is nonzero
by an exponentially distributed r.v.~with rate $\lambda^{-1}$ where $\lambda = 0.125, 0.5, 2.5$. 
Modeling the amount of 
rainfall with this r.v.~allows us to assess $\TDE$s when the 
$\SNR$ varies from high to low. Indeed, since the variance of such r.v.~is $\lambda^{-2}$ and
the quadratic variation of the impulse function $f$ is a constant, $\kappa$ say, 
$\SNR = \kappa / \lambda$.

\subsection{Discussion of simulation results}~\label{subsec.Results}

Our findings are summarized in Table~\ref{tab_mean_sd_ExpRainfall}. 
For $\tau=37, 110$ the grid search was done over 40\% of $\mc G_n$, 
see Section~\ref{sec.methods}, whereas for $\tau=183$ we used 50\%.
For a high $\SNR$ ($\lambda=0.125$), as expected, any estimator works appropriately independently
of delay and eco-region.

\subsubsection{Cases $\tau=37, 110$}
For a moderate $\SNR$ ($\lambda=0.5$) and for a short
delay ($\tau=37$), the trimmed estimators behave poorly as they exhibit a large variance,
regardless of these estimators, the LASSO-based estimators seem to outperform the Pearson-based
across the eco-regions. As the delay increases ($\tau=110$) the variance of the $\TDE$s
seems to reduce and now $\LASSOCVCor$ is the best in
Madrense, Mezquital and Plateau Plains whereas for Interior Plains the
unscaled estimator $\LASSO{0.1}$ is the best option.

In the difficult, yet interesting, case of low $\SNR$ ($\lambda = 2.5$) once again the trimmed
estimators exhibit poor accuracy and large variability when the delay is short; in this
case $\LASSOCV$ seems to outperform all the other estimators. As in the paragraph above
accuracy and variability improve as the delay increases and $\LASSOCVCor$
is the best option across the eco-regions.

\subsubsection{Case $\tau=183$}
For a moderate $\SNR$ and across all eco-regions any estimator seems 
appropriate although $\Pn$ and $\LASSOCV$ are marginally better. For a low $\SNR$ it seems
that the best accuracy and smallest variance is achieved by the unscaled estimators
$\Pn$, $\LASSO{0.1}$ and $\LASSOCV$.

\section{Applications}~\label{sec.apps}


We have applied the estimators introduced in Section~\ref{sec.sims} to the time series presented
in Section~\ref{sec.intro}; we excluded the 2008 images from the analysis due to their poor quality;
grid search was done over $\pm 3$ months.

\begin{figure}[htb]
\centering
    \scalebox{.75}{{\includegraphics[width=\linewidth]{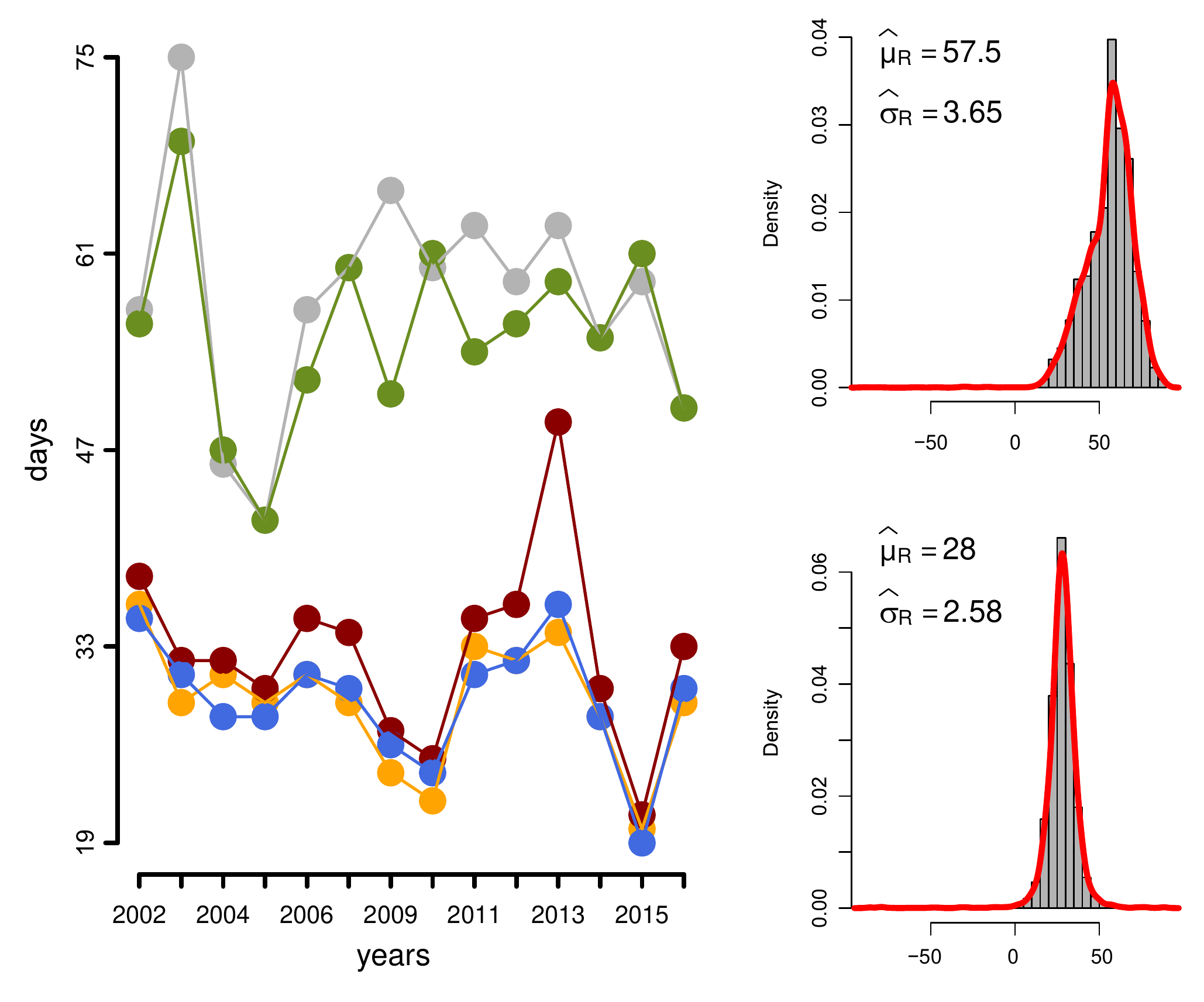}}}
    


    \vspace{-.125cm}
    \caption{\footnotesize{Mezquital eco-region. Left: Median of $\TDE$s (\circleCol{gray70}) $\Pn$,
    (\circleCol{orangeR}) $\PnCor$,
    (\circleCol{red4}) $\PnStd$, (\circleCol{olivedrab}) $\LASSOCV$, (\circleCol{royalblue}) $\LASSOCVCor$.
    Top-right: Estimated density of $\Pn$ from 2002 to 2016 .
    Bottom-right: Estimated density of $\LASSOCVCor$  from 2002 to 2016.
    \label{Fig2}}}
\end{figure}

In addition to getting the optimal lag between daily precipitation and NDVI by year, 
via \eqref{eq.TiDE}-\eqref{eq.sTiDE}, we calculated the $p$-value associated to the null 
hypothesis of \emph{no correlation} between these variables at the optimal lag. We ended up
with 14 annual maps of significant delay estimates which are
summarized, via the median (a well-known robust statistics), and shown in Fig.\protect\ref{Fig1B}
and \protect\ref{Fig1C} for the $\Pn$ and $\LASSOCVCor$ estimators, respectively. 
There are marked spatial differences between these estimators. For instance, 
their median and (a robust) s.d. estimate are different
over the Mezquital eco-region, 28 (2.57) for $\LASSOCVCor$ and 58 (3.65) for $\Pn$.
Observe also, Fig.\protect\ref{Fig2}, the narrow that $\LASSOCVCor$'s distribution is about its center which
contrast with the dispersion of $\Pn$'s distribution; their time dynamics is also different.

We show our analysis on Mezquital eco-region only but the differences between unscaled and trimmed $\TDE$s 
were found also in the other three eco-regions, see Supplementary Materials~\ref{sec.supp}.

\section{Conclusion}~\label{sec.conclusion}
By incorporating sparsity we introduced a simple method to estimate the time delay between
precipitation and NDVI satellite imagery time series aiming to reduce the variance 
of the canonical Pearson-based estimator. We showed that this goal was achieved
first in simulations (and varying $\SNR$) and then in applications (across regions with different ecological characteristics).
In practice, besides its variance reduction, the decision of (perhaps) favor the LASSO-based estimators 
over the canonical one should be weighted with prior information (more biophysical variables) 
on the characteristics of the region of interest;
the interaction of more variables to describe time relationships requires
further investigation.

\section{Acknowledgements}~\label{sec.acknow}
The author would like to thank Rene Colditz with the Climate Action at the European Commission for introducing him
to the topic of time relationships. This paper benefited also from helpful discussions with 
Carlos Troche, Isabel Cruz and Gabriela Villamil at CONABIO and Irina Gaynanova (Department of Statistics, Texas A\&M University)
and Michael Grabchak (Department of Mathematics and Statistics, UNC Charlotte).

\bibliographystyle{apalike} 
\bibliography{TiDEBib}
\newpage
\section{Supplementary Materials}~\label{sec.supp}

\begin{table}[htb]
\centering
\scalebox{0.65}{
\begin{tabular}{R Q Q Q P Q Q Q P Q Q Q} 
  \toprule[1.25pt]
      
  $\TDE$s & \multicolumn{3}{c}{$\tau=37$} && \multicolumn{3}{c}{$\tau=110$} && \multicolumn{3}{c}{$\tau=183$} \\ 
          & \multicolumn{3}{c}{Exp($\lambda$) errors} && \multicolumn{3}{c}{Exp($\lambda$) errors} && \multicolumn{3}{c}{Exp($\lambda$) errors}\\
    
  \midrule[1.25pt] 
  & \multicolumn{1}{c}{$\lambda = 0.125$} & $\lambda = 0.5$ & $\lambda = 2.5$ 
  &
  & \multicolumn{1}{c}{$\lambda = 0.125$} & $\lambda = 0.5$ & $\lambda = 2.5$ 
  &
  & \multicolumn{1}{c}{$\lambda = 0.125$} & $\lambda = 0.5$ & $\lambda = 2.5$ \\
  \cmidrule[1.25pt]{2-4}
  \cmidrule[1.25pt]{6-8}
  \cmidrule[1.25pt]{10-12} 
  
  
  \multicolumn{12}{c}{Madrense}\\
  
  $\Pn$ & 0.001 & 0.286 & 14.754 && 0.001 & 0.101 & 5.495 && 0.000 & 0.002 & 0.063 \\ 
  $\PnCor$ & 0.000 & 106.998 & 360.854 && 0.000 & 0.023 & 1.782 && 0.000 & 0.008 & 346.795 \\ 
  $\PnStd$ & 0.000 & 0.160 & 31.066 && 0.000 & 0.040 & 1.966 && 0.000 & 0.004 & 0.218 \\ 
  $\LASSO{0.1}$ & 0.000 & 0.121 & 14.610 && 0.000 & 0.039 & 5.182 && 0.000 & 0.004 & 0.062 \\ 
  $\LASSOCor{0.1}$ & 0.000 & 0.036 & 1398.367 && 0.000 & 0.013 & 1.384 && 0.000 & 0.008 & 0.780 \\ 
  $\LASSOCV$ & 0.000 & 0.257 & 13.349 && 0.001 & 0.101 & 4.999 && 0.000 & 0.002 & 0.078 \\ 
  $\LASSOCVCor$ & 0.000 & 35.684 & 224.517 && 0.000 & 0.013 & 1.148 && 0.000 & 0.007 & 0.684 \\ 
  
  \multicolumn{12}{c}{Mezquital}\\

    $\Pn$ & 0.000 & 0.047 & 13.025 && 0.003 & 0.479 & 13.563 && 0.000 & 0.002 & 0.118 \\ 
  $\PnCor$ & 0.000 & 522.814 & 1265.681 && 0.000 & 0.082 & 6.124 && 0.000 & 0.034 & 126.105 \\ 
  $\PnStd$ & 0.000 & 0.043 & 290.854 && 0.000 & 0.114 & 6.420 && 0.000 & 0.009 & 0.571 \\ 
  $\LASSO{0.1}$ & 0.000 & 0.051 & 12.970 && 0.000 & 0.089 & 13.101 && 0.000 & 0.006 & 0.120 \\ 
  $\LASSOCor{0.1}$ & 0.000 & 0.051 & 3292.637 && 0.000 & 0.057 & 4.129 && 0.000 & 0.031 & 1.289 \\ 
  $\LASSOCV$ & 0.000 & 0.043 & 8.383 && 0.003 & 0.443 & 11.348 && 0.000 & 0.003 & 0.125 \\ 
  $\LASSOCVCor$ & 0.000 & 59.239 & 984.263 && 0.000 & 0.051 & 2.820 && 0.000 & 0.027 & 1.209 \\ 

  \multicolumn{12}{c}{Interior Plains}\\

    $\Pn$ & 0.000 & 0.140 & 27.744 && 0.001 & 0.492 & 16.470 && 0.000 & 0.008 & 0.821 \\ 
  $\PnCor$ & 0.000 & 831.560 & 2134.619 && 0.000 & 0.176 & 372.674 && 0.001 & 0.185 & 1001.198 \\ 
  $\PnStd$ & 0.000 & 0.076 & 624.842 && 0.000 & 0.099 & 7.313 && 0.001 & 0.030 & 3.476 \\ 
  $\LASSO{0.1}$ & 0.000 & 0.057 & 25.079 && 0.000 & 0.092 & 16.156 && 0.001 & 0.021 & 0.769 \\ 
  $\LASSOCor{0.1}$ & 0.000 & 0.122 & 4324.318 && 0.000 & 0.137 & 271.041 && 0.001 & 0.124 & 98.804 \\ 
  $\LASSOCV$ & 0.000 & 0.126 & 23.340 && 0.000 & 0.458 & 12.482 && 0.000 & 0.011 & 0.806 \\ 
  $\LASSOCVCor$ & 0.000 & 83.054 & 1617.145 && 0.000 & 0.128 & 6.464 && 0.001 & 0.122 & 52.359 \\ 

  \multicolumn{12}{c}{Plateau Plains}\\
  
  $\Pn$ & 0.000 & 0.445 & 21.739 && 0.001 & 0.302 & 14.249 && 0.000 & 0.004 & 0.981 \\ 
  $\PnCor$ & 0.000 & 558.475 & 1273.985 && 0.000 & 0.078 & 9.115 && 0.000 & 0.090 & 40.032 \\ 
  $\PnStd$ & 0.000 & 0.061 & 438.903 && 0.000 & 0.082 & 8.661 && 0.000 & 0.044 & 2.211 \\ 
  $\LASSO{0.1}$ & 0.000 & 0.057 & 21.171 && 0.000 & 0.090 & 14.565 && 0.000 & 0.009 & 0.957 \\ 
  $\LASSOCor{0.1}$ & 0.000 & 0.075 & 3644.262 && 0.000 & 0.053 & 36.607 && 0.000 & 0.071 & 3.061 \\ 
  $\LASSOCV$ & 0.000 & 0.439 & 19.910 && 0.001 & 0.257 & 12.025 && 0.000 & 0.004 & 0.958 \\ 
  $\LASSOCVCor$ & 0.000 & 47.604 & 1080.126 && 0.000 & 0.052 & 36.040 && 0.000 & 0.070 & 2.799 \\ 

  \bottomrule[1.25pt]
\end{tabular}}
\caption{\small{Mean squared error corresponding to estimates shown in Table~\ref{tab_mean_sd_ExpRainfall}.}~ \label{tab_mse_ExpRainfall}}
\end{table}

\begin{figure}[htb]
\centering
    \scalebox{1}{{\includegraphics[width=\linewidth]{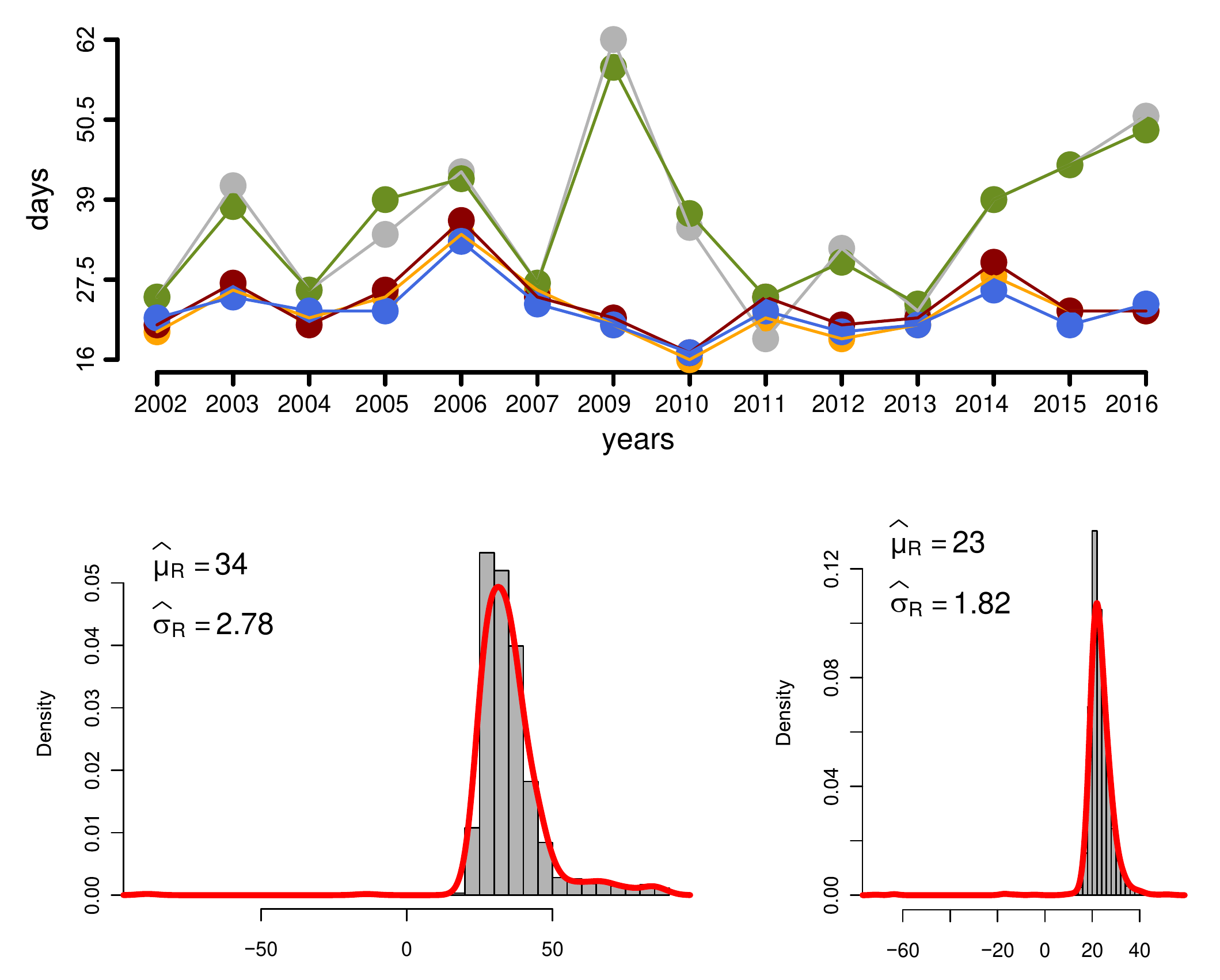}}}
    \vspace{-.125cm}
    \caption{\footnotesize{Madrense eco-region. Left: Median of $\TDE$s 
    (\circleCol{gray70}) $\Pn$, (\circleCol{orangeR}) $\PnCor$,
    (\circleCol{red4}) $\PnStd$, (\circleCol{olivedrab}) $\LASSOCV$, (\circleCol{royalblue}) $\LASSOCVCor$.
    Top-right: Estimated density of $\Pn$ from 2002 to 2016 .
    Bottom-right: Estimated density of $\LASSOCVCor$  from 2002 to 2016.\label{Fig3}}}
\end{figure}

\begin{figure}[htb]
\centering
    \scalebox{1}{{\includegraphics[width=\linewidth]{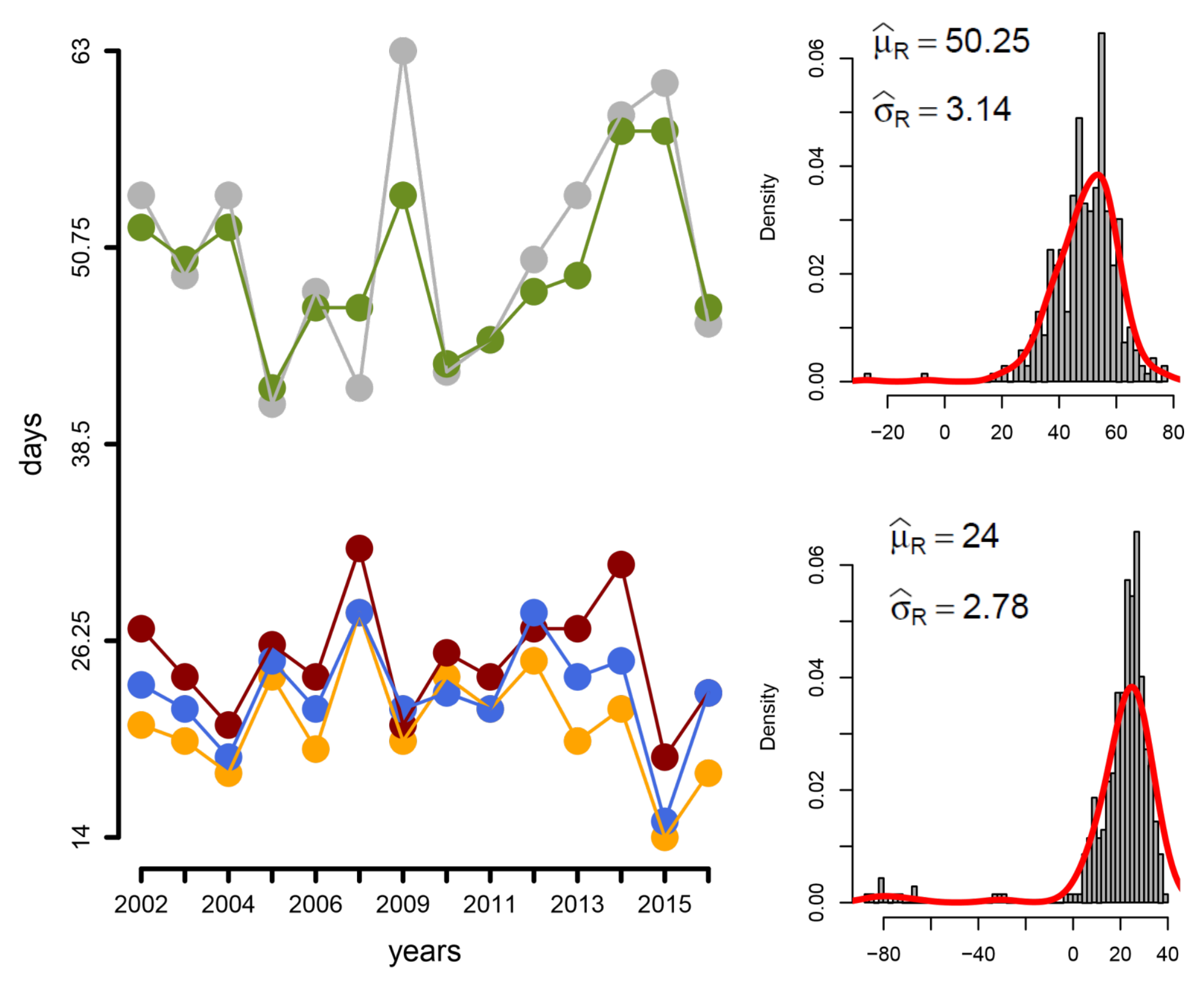}}}
    \vspace{-.125cm}
    \caption{\footnotesize{Interior Plains eco-region. Left: Median of $\TDE$s 
    (\circleCol{gray70}) $\Pn$, (\circleCol{orangeR}) $\PnCor$,
    (\circleCol{red4}) $\PnStd$, (\circleCol{olivedrab}) $\LASSOCV$, (\circleCol{royalblue}) $\LASSOCVCor$.
    Top-right: Estimated density of $\Pn$ from 2002 to 2016 .
    Bottom-right: Estimated density of $\LASSOCVCor$  from 2002 to 2016.\label{Fig4}}}
\end{figure}

\begin{figure}[htb]
\centering
    \scalebox{1}{{\includegraphics[width=\linewidth]{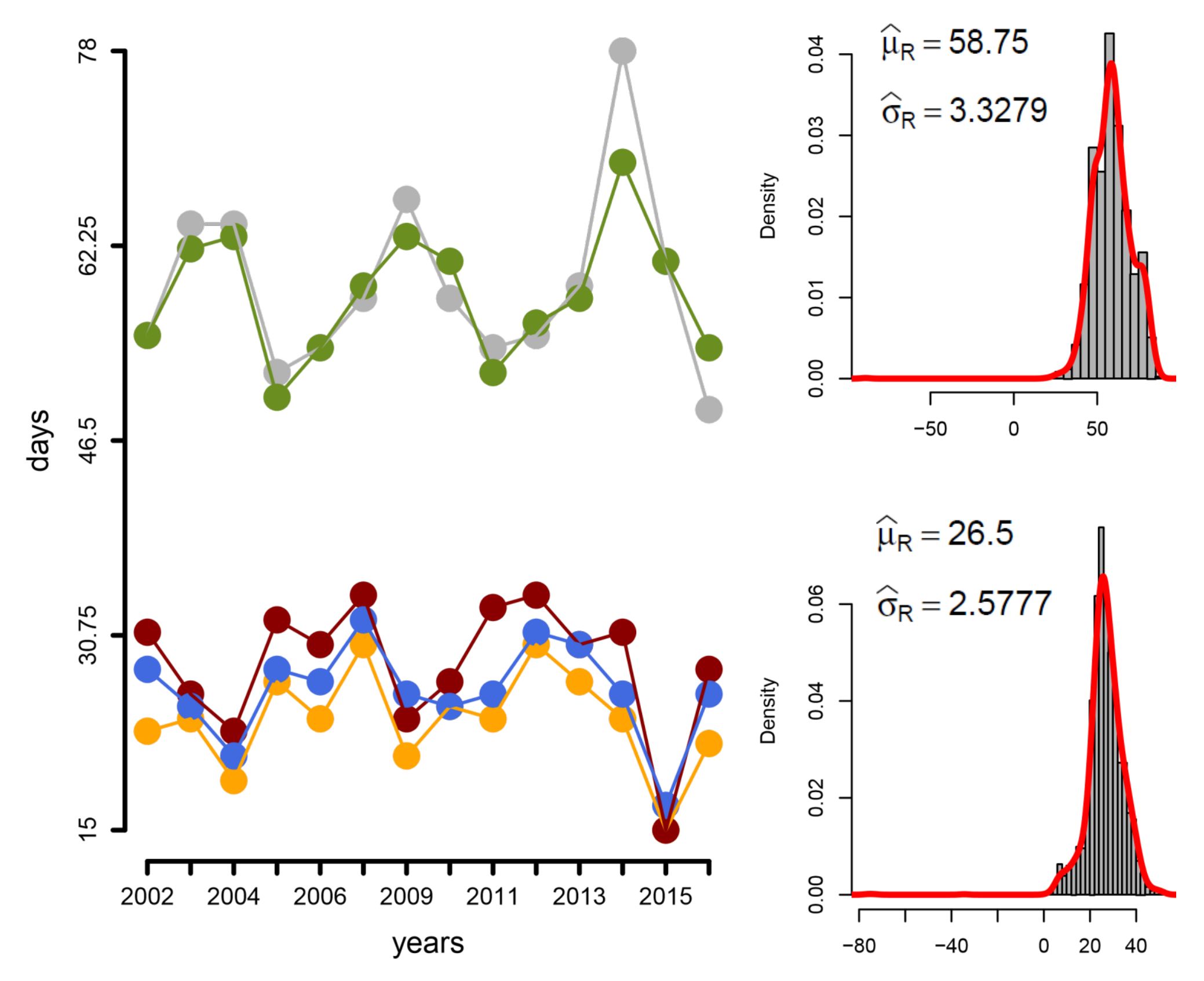}}}
    \vspace{-.125cm}
    \caption{\footnotesize{Plateau Plains eco-region. Left: Median of $\TDE$s 
    (\circleCol{gray70}) $\Pn$, (\circleCol{orangeR}) $\PnCor$,
    (\circleCol{red4}) $\PnStd$, (\circleCol{olivedrab}) $\LASSOCV$, (\circleCol{royalblue}) $\LASSOCVCor$.
    Top-right: Estimated density of $\Pn$ from 2002 to 2016 .
    Bottom-right: Estimated density of $\LASSOCVCor$  from 2002 to 2016.\label{Fig5}}}
\end{figure}

\end{document}